\documentclass[11pt]{scrartcl}

\usepackage{amsmath}
\usepackage{upgreek}
\usepackage{graphicx}
\usepackage{cite}
\begin{document}

\title{Elastic measurements on macroscopic three-dimensional pentamode metamaterials}
\author{\normalsize{Robert~Schittny$^1$, Tiemo~B\"{u}ckmann$^1$, Muamer~Kadic$^1$, and Martin~Wegener$^{1,2}$}}
\date{}
\maketitle

\begin{center}
\textit{\noindent$^1$Institute of Applied Physics, Karlsruhe Institute of Technology (KIT), 76128~Karlsruhe, Germany}

\textit{\noindent$^2$Institute of Nanotechnology, Karlsruhe Institute of Technology (KIT), 76128~Karlsruhe, Germany}
\end{center}

\begin{abstract}
Pentamode metamaterials approximate tailorable artificial liquids. Recently, microscopic versions of
these intricate three-dimensional structures have been fabricated, but direct
experimental characterization has not been possible yet. Here, using three-dimensional printing, we
fabricate macroscopic polymer-based samples with many different combinations of the small connection
diameter $d$ and the lattice constant $a$. Direct measurements of the static shear modulus and the
Young's modulus reveal that both scale approximately according to $(d/a)^3$, in good agreement with
continuum-mechanics calculations. For the smallest accessible values of $d/a\approx 1.5\%$, we find 
derived ratios of bulk modulus $B$ to shear modulus $G$ of $B/G \approx 1000$.
\end{abstract}

Pentamode elastic metamaterials\cite{Milton1995,Kadic2012,Martin2012} are rationally designed artificial crystals allowing for obtaining
an effective metamaterial shear modulus $G$ which can be orders of magnitude smaller than the
effective bulk modulus $B$. Intuitively, this means that it is much easier to change the
shape of such a metamaterial (while fixing its volume) than it is to change its volume (while fixing
its shape). In this sense, pentamode metamaterials approximate the elastic properties of liquids. In
contrast to liquids, however, it is straightforward to envision stable and intentionally spatially
inhomogeneous as well as anisotropic \cite{Kadic2013,Layman2013} pentamode metamaterial architectures. Normal
liquids would simply flow away or intermix. Hence, such metamaterial structures are promising for
experimentally translating the concepts of transformation optics \cite{Pendry2006,Schurig2006,Leonhardt2010} to
elastostatics and elastodynamics \cite{Milton2006,Norris2008,Brun2009,Norris2011}. For example, one could design and
realize elastostatic cloaks, which would make some hard interior unfeelable from the outside.

Pentamode metamaterials were already suggested theoretically in 1995\cite{Milton1995}, but have
been realized experimentally only in 2012\cite{Kadic2012}. However, the microscopic samples
fabricated there \cite{Kadic2012} by using dip-in direct laser writing could not be characterized
directly. In this Letter, taking advantage of the scalability of continuum mechanics, we fabricate
much larger macroscopic versions of pentamode metamaterials with lattice constants of $a\approx
1\,\text{cm}$. This step enables direct measurement of the elastic properties as a function of the
critical geometrical parameter, i.e., the diameter of the small connections within the pentamode
unit cell.

\begin{figure}
\centering
\includegraphics{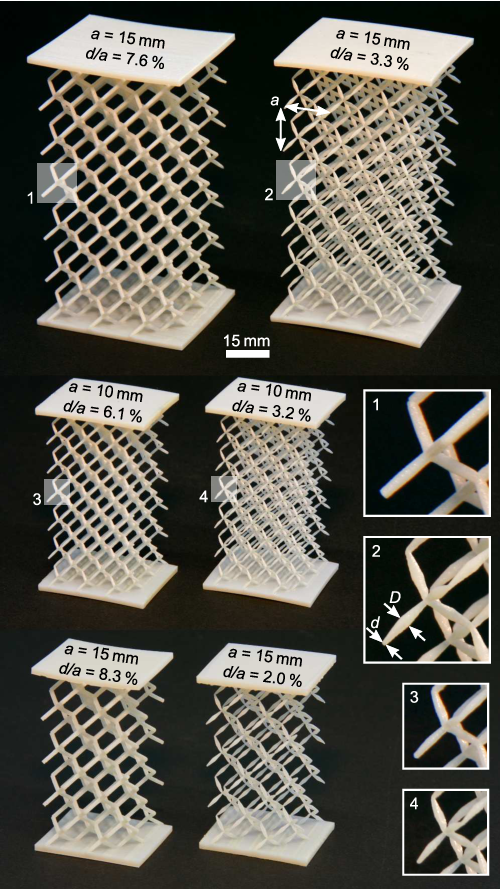}
\caption{Gallery of photographs of selected macroscopic polymer-based pentamode metamaterial samples
fabricated using three-dimensional printing. The insets show magnified parts of the samples as indicated
by the white overlays.}
\label{fig1_samples}
\end{figure}

Fig.~\ref{fig1_samples} shows a gallery of selected photographs of a number of pentamode
metamaterial structures that we have fabricated using a three-dimensional printer (Objet30 by
formerly Objet, now Stratasys, USA).  For the printing process of the geometrically complex
pentamode metamaterials, a support material is needed that holds all overhanging parts and
can be removed later. We choose Stratasys' proprietary ``FullCure850 VeroGray'' polymer as the model
material and pure ``FullCure705 Support'' as support, which can be etched out after printing in a
bath of NaOH base. The default support material (a fine-grained mixture of ``FullCure850 VeroGray''
and ``FullCure705 Support'') has proven to be impossible to remove from the delicate pentamode
structures.
The pentamode metamaterials are composed of double-cone elements, the tips of which touch each other
at a set of fictitious points forming a diamond lattice \cite{Milton1995,Kadic2012}. We have
previously shown by numerical modeling \cite{Kadic2012,Martin2012} that the diameter $d$ of the
resulting touching region compared to the lattice constant $a$ of the face-centered-cubic (fcc)
translational lattice (see Fig.~\ref{fig1_samples} for the geometrical dimensions) is the most
important parameter in terms of influence on the effective properties. Precisely, the ratio of 
bulk to shear modulus scales like $B/G\propto
(a/d)^2$. This scaling results from the approximate dependencies $B\propto d/a$ and $G \propto
(d/a)^3$. In contrast, we have shown that the diameter $D$ of the thick end of the cones is less
important \cite{Kadic2012}. Intuitively, as so often in life, ``the weakest link in a chain
determines its properties''.

The relevant $B/G$ ratio is directly connected to the effective metamaterial Poisson's ratio
$\nu$\cite{Gould1994}, which therefore could in principle be used as a measure for characterization. For
$B/G\rightarrow \infty$, one gets $\nu \approx 0.5(1-(B/G)^{-1}) \rightarrow 0.5$\cite{Kadic2012}.
This means that very large $B/G$ ratios would have to be determined from minute deviations of $\nu$
from its upper bound of 0.5, requiring measurements of $\nu$ with at least three significant digits.
Such precision measurements appear neither attractive nor in reach to us. Thus, we rather measure
the shear modulus via its definition, i.e., we impose a shear displacement and measure the shear
force using a force cell (``K3D60N 10\,N'' from ME-Me{\ss}systeme, Germany). Together with the known
area the force is acting upon, the shear modulus follows directly and quantitatively.

\begin{figure}
\centering
\includegraphics{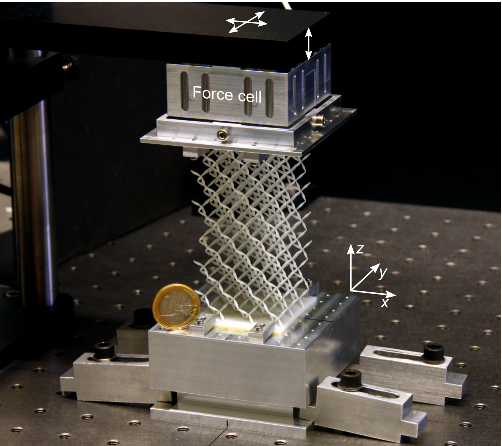}
\caption{Photograph of the experimental setup including a force cell, allowing for direct shear and
Young's modulus measurements. The bottom and top metal stamps of this setup are rigidly attached to
polymer plates that have been fabricated monolithically in the three-dimensional polymer-printing
process.}
\label{fig2_setup}
\end{figure}

\begin{figure}
\centering
\includegraphics{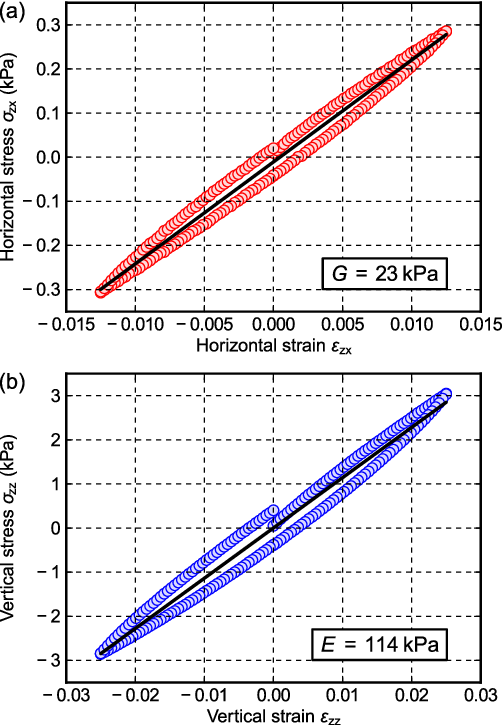}
\caption{(a) Example of an individual shear-modulus measurement (red circles). (b) Example of one Young's-modulus
measurement (blue circles). The moduli are obtained from the slopes of the fitted straight
lines. The geometrical parameters (see Fig.~\ref{fig1_samples}) of the sample
composed of $2 \times 2 \times 4$ extended unit cells are: $d = 0.71\,\text{mm}$, $D = 1.32\,\text{mm}$,
and $a = 15\,\text{mm}$.}
\label{fig3_meas}
\end{figure}

\begin{figure}
\centering
\includegraphics{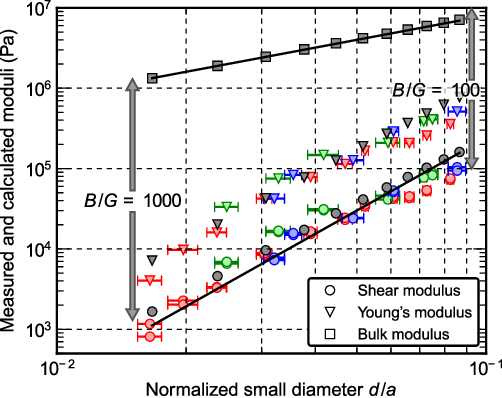}
\caption{Elastic moduli (real parts) versus the ratio $d/$a of the diameter of the small double-cone
connections $d$ and the pentamode fcc lattice constant $a$ in double-logarithmic representation.
Shown are the shear modulus $G$ as circles, the Young's modulus $E$ as triangles, and the bulk
modulus $B$ as squares. The colored symbols correspond to experimental data (also see
Fig.~\ref{fig3_meas}), the black symbols to numerically calculated values for the geometrical
parameters taken from the experiment. The sample parameters are $a = 1.5\,\text{cm}$ and $2 \times 2
\times 4$ extended unit cells (red), $a = 1.5\,\text{cm}$ and $3 \times 3 \times 6$ extended unit
cells (green), and $a = 1.0\,\text{cm}$ and $3 \times 3 \times 6$ extended unit cells (blue). The
only other geometrical parameter, the diameter of the thick end of the double cones is fixed to
$D/a=0.087$ for all samples. The two black solid straight lines are guides to the eye (not fits) and
correspond to scalings of $G \propto (d/a)^3$ and $B \propto (d/a)^1$, respectively. The two
vertical double-arrows on the left and on the right highlight the approximate $B/G$ ratio for the 
smallest and the largest investigated ratios of $d/a$. A ratio of $B/G = 1000$ means that the pentamode
metamaterial approximates well the properties of an isotropic liquid.}
\label{fig4_summary}
\end{figure}

A photograph of the corresponding computer-controlled measurement setup is shown in
Fig.~\ref{fig2_setup}. To obtain well-defined and reproducible boundary conditions, we directly and
monolithically print each of the to-be-characterized pentamode metamaterial samples with rigid
plates on the top and bottom which then are fixed to the metal stamps of the measurement setup.  A
representative force versus strain measurement within the linear regime is depicted in
Fig.~\ref{fig3_meas}(a). As usual for viscoelastic materials, one gets a certain hysteresis when
linearly ramping up and down the shear strain over a total time period of about 100\,s. 
The purely elastic component of the shear
modulus is obtained from the fitted slope (see straight line in Fig.~\ref{fig3_meas}(a)). Repeating
measurements like that for many different samples with different values of $d$ and $a$ and for both
the $x$- and the $y$-direction leads to the data points summarized in Fig.~\ref{fig4_summary}, which
uses a double-logarithmic representation of the moduli in dependence on $d/a$. Apart from some scatter,
the measured data points (colored circles) roughly follow the anticipated scaling $G\propto (d/a)^3$
(see black straight line). We note that the geometrical parameters $d$, $D$, and $a$ have directly
been determined from an analysis of photographs of the actually fabricated and investigated samples.
The measured geometrical parameters for $d$ and $D$ show deviations from the nominal ones (that are fed into the
3D printer) by approximately 10\%. The horizontal error bars in Fig.~\ref{fig4_summary} correspond
to the scatter of the determined $d$ values in each sample. The measurements of $G$ (colored circles) are
compared with numerical calculations for the shear modulus (black circles). For details on the
numerical calculations, we refer the reader to Ref.~\cite{Kadic2012}. We take the identical
geometrical parameters and the identical number of extended pentamode unit cells as in the present
experiments. Using a Young's modulus of 1.4\,GPa and a Poisson's ratio of 0.4 (the latter being not
critical at all \cite{Kadic2012}) for the constituent polymer material, we obtain good agreement
with the experimental data (compare black and colored circles in Fig.~\ref{fig4_summary}). Here, the
Young's modulus of 1.4\,GPa has been the only adjustable parameter. According to the underlying
continuum-mechanics equations, it only scales all effective metamaterial moduli by a common factor,
but it strictly neither influences the exponent versus $d/a$ nor any ratio between effective
metamaterial moduli. We have also performed different independent Young's modulus measurements on
bulk pieces of the printed constituent polymer ``FullCure850 VeroGray'', leading to values of
0.7--2.0\,GPa. The value quoted by the provider is 2--3\,GPa. The value of 1.4\,GPa chosen above
lies within this significant scatter.

Next, one would ideally like to measure the effective metamaterial bulk modulus $B$ directly as
well. To do so, in principle, one needs to simultaneously and equally push from all six sides of the
cuboid samples under sliding boundary conditions 
and measure the
corresponding displacements and forces. Such a measurement appears challenging to us. We thus rather
follow a different, somewhat more indirect but readily feasible approach: We also measure the
Young's modulus, again compare to theory, and then, assuming good overall agreement, use theory to
derive the bulk modulus. In more detail, it is simple to show that the effective metamaterial
Young's modulus $E$ should follow $E \approx 3G$ in the limit of $B/G \rightarrow \infty$. To
measure the Young's modulus, we again use the setup shown in Fig.~\ref{fig2_setup} but impose a
displacement along the vertical direction (i.e., a given strain) and measure the resulting vertical
force component with the force cell. From these, the Young's modulus follows according to its
definition.  To allow for direct comparison with the shear data, we use the same samples as for the
shear measurements. Due to the mentioned mounting plates, however, the samples experience fixed
boundary conditions at the stamps along both horizontal directions. In an ideal Young's modulus
measurement, the boundary conditions should rather be sliding. Thus, we use fixed boundary
conditions in our numerical modeling as well. An example of a measured stress-strain relation is
depicted in Fig.~\ref{fig3_meas}(b). The Young's modulus results from the slope of the indicated
straight line.  Repeating this experiment for all samples leads to the set of colored triangles
exhibited in Fig.~\ref{fig4_summary}. The corresponding calculated values are also shown there as
black triangles. Obviously, without any further adjustable parameters, experiment and theory agree
nicely in regard to the Young's modulus as well. In the double-logarithmic representation, the
Young's modulus data points are parallel to the shear modulus data points. This means that the
Young's modulus scales with the same exponent of 3 versus $d/a$. Based on the good overall agreement
between experiment and theory in regard to $G$ and $E$ versus $d/a$, we consider calculated
values of $B$ as reliable and trustworthy. Since edge effects would dominate in this calculation for 
the sample sizes used here, we rather consider a fictitious infinitely extended crystal. We assume that
all extended fcc unit cells therein are compressed equally in all three cubic directions upon imposing a pressure 
from these three directions simultaneously. Technically, choosing a coordinate system with a fixed
center of gravity, this translates into ``anti-symmetric'' boundary conditions for the normal 
component of the displacement vector on opposing faces of the extended fcc unit cell.
The calculated data points for $B$ are shown as black squares in Fig.~\ref{fig4_summary}. At the smallest accessible
ratios of $d/a \approx 1.5\%$ on the left-hand side, the $B/G$ ratio is as large as 1000 (see
double-arrow). Even at the largest ratios of $d/a \approx 9\%$ on the right-hand side (where the
structure's double-cones essentially become straight rods), the $B/G$ ratio is still on the order 
of 100 (see double-arrow).

Let us finish by noting that our discussion above has tacitly assumed that the effective
metamaterial properties are isotropic (otherwise $B$ and $G$ would turn from scalars into tensors),
while the pentamode unit cell only guarantees cubic crystal symmetry. In general, for a cubic
crystal, Poisson's ratio is not isotropic and can even exhibit rather complex and pronounced
direction dependencies \cite{Wojciechowski2005}. However, it has been shown\cite{Wojciechowski2005} 
that a Poisson's ratio approaching $0.5$ from below for pushing along the principal axes of a cubic
crystal leads to an isotropic Poisson's ratio $\nu \rightarrow 0.5$. This means
that our metamaterial structures approximate isotropic liquids. Anisotropic versions of
three-dimensional \cite{Kadic2013} and two-dimensional \cite{Layman2013} pentamode metamaterials
have also been discussed theoretically (strictly speaking, there is no such thing as a
two-dimensional isotropic pentamode metamaterial, it is rather a bimode metamaterial). It is
presently not clear though how to fully characterize the complex elasticity tensor of such 
structures with non-cubic crystal symmetry in direct static experiments.

In conclusion, we have measured the effective shear modulus and the effective Young's modulus on a
large set of different macroscopic three-dimensional polymer-based pentamode metamaterials. The good
agreement of experiment and theory allows us to extract the ratios of bulk to shear modulus. We find
ratios as large as $B/G \approx 1000$, which were previously predicted only purely theoretically
\cite{Kadic2012}. This finding raises hopes that three-dimensional elastic cloaks are indeed in
reach experimentally. For example, theoretical work on pentamode-based circular cloaks
\cite{Urzhumov2010} has shown good cloaking for ratios of $B/G$ of 100 and even excellent cloaking
for ratios of 1000. A bulk of theoretical work which has assumed that ideal pentamode materials are
available \cite{Norris2009,Urzhumov2010,Scandrett2010,Cipolla2011,Gokhale2012} is now backed up by 
an experimental characterization of three-dimensional pentamode elastic metamaterials in the 
static limit.

We thank Prof. Manfred Wilhelm (KIT) for rheological measurements on the bulk printing material. 
We acknowledge support from the DFG-Center for Functional Nanostructures (CFN) and the Karlsruhe 
School of Optics \& Photonics (KSOP).

\end{document}